\documentclass{article}
\usepackage{amsfonts}
\usepackage{amssymb}
\usepackage{amsmath}
\usepackage{amsthm}
\usepackage[english]{babel}
\begin{document}
\sloppy

\title{\bf Smooth Horizons and Quantum Ripples}

\author{{\bf Alexey Golovnev}\\
{\small {\it Saint Petersburg State University, high energy physics department,}}\\
{\small \it Ulyanovskaya ul., d. 1; 198504 Saint-Petersburg, Petrodvoretz; Russia}\\
{\small agolovnev@yandex.ru, \quad golovnev@hep.phys.spbu.ru}}
\date{}

\twocolumn[\maketitle
\begin{center} {\bf Abstract} \end{center}

Black Holes are unique objects which allow for meaningful theoretical studies of strong gravity and
even quantum gravity effects. An infalling and a distant observer would have very different views on the structure of the world. However, a careful analysis has shown that it entails no genuine contradictions for physics, and the paradigm of observer complementarity has been coined. Recently this picture was put into doubt. In particular, it was argued that in old Black Holes a firewall must form in order to protect the basic principles of quantum mechanics. This AMPS paradox has already been discussed in a vast number of papers with different attitudes and conclusions. Here we want to argue that a possible source of confusion is neglection of quantum gravity effects. Contrary to widespread perception, it does not necessarily mean that effective field theory is inapplicable in rather smooth neighbourhoods of large Black Hole horizons. The real offender might be an attempt to consistently use it over the huge distances from the near-horizon zone of old Black Holes to the early radiation. We give simple estimates to support this viewpoint and show how the Page time and (somewhat more speculative) scrambling time do appear.

\vspace{1cm}]

It is an amazing fact about Black Holes that they can emit particles \cite{Hawking}. Assuming that this radiation is purely thermal and that a Black Hole will eventually evaporate completely, the information which has been swallowed by the Black Hole during its lifetime is lost, in contradiction to the unitary nature of quantum mechanics. It was then established that unitarity can still be saved if an effective description of the outer near-horizon region is allowed in terms of a Planck-width stretched horizon which can absorb, thermalise and emit information \cite{complementarity}. Of course, due to equivalence principle, an infalling observer should not experience anything special while crossing the horizon. However, there would be no way for him to share this knowledge with the distant colleague, and therefore one can avoid running into a contradiction by adopting the fancy viewpoint that there is no need for a universal global effective field theory description of physics in the whole spacetime. It marked the birth of the Black Hole complementarity paradigm \cite{complementarity}.

\vspace{0.5cm}

In order to better understand the relevant time scales of evaporation, recall that a Black Hole of mass $M$ has Hawking temperature $${\mathrm k}_B T=\frac{\hbar c^3}{8\pi G M}$$ with the usual notations for the Planck constant $\hbar\equiv\frac{h}{2\pi}$, the speed of light $c$, the gravitational constant $G$, and the Boltzmann constant ${\mathrm k}_B$. Under the Stefan-Boltzmann law, the emissive power scales as $T^4\propto \frac{1}{M^4}$ while the horizon area $\propto M^2$ which implies $\dot M\propto \frac{1}{M^2}$ for the rate of energy loss, and therefore the characteristic time of evaporation grows with mass as $M^3$ \cite{Page}, or more explicitly $t\sim \left(\frac{M}{m_{Pl}}\right)^3 t_{Pl}$ where the subscript $Pl$ stands for the Planckian quantities.

\vspace{0.5cm}

The problem \cite{AMPS} presents itself when the Black Hole has emitted one half of its total entropy\footnote{Note that, under the name of energy curtains, this paradox was earlier reported in Ref. \cite{Braunstein}, see also Refs. \cite{Braunstein2,Braunstein3}.}, or at the Page time $$t_{\mathrm P}\sim \left(\frac{M}{m_{Pl}}\right)^3\cdot t_{Pl}.$$ In this case the already emitted radiation contains practically all information which has gone into the Black Hole \cite{entropyPage}, and therefore must be fully entangled with the late radiation appearing in the near-horizon zone. On the other hand, from the viewpoint of an infalling observer, both sides of the horizon are just the two halves of very flat (almost Minkowski) space. Under the spell of equivalence principle, the freely falling observer has all the legal rights to expect meeting the vacuum state there. And the vacuum state is a very regular thing characterised by maximal entanglement of those two halves. Of course, being simultaneously maximally entangled with two different systems is impossible. The corresponding general principle of quantum mechanics is known under the beautiful name of monogamy of entanglement.

There is apparently a strong tension between the equivalence principle and quantum mechanics which becomes evident at the Page time. But it was also argued \cite{AMPS} that the same should be true much earlier, at the scrambling time when the information of constituent matter has already been scrambled (thermalised) by the Black Hole. This time scale is much smaller. Actually, there are some reasonably good reasons to believe that it must be of order  $$t_{scr}\sim \left(\frac{M}{m_{Pl}}\log \frac{M}{m_{Pl}}\right)\cdot t_{Pl}$$ which makes the Black Holes amazingly fast scramblers \cite{fastscrambler}. The status of this time scale is however unclear \cite{Susskind1}. 

\vspace{0.5cm}

With one time scale or another, the way out proposed in Ref. \cite{AMPS} was a firewall just behind the horizon. In other words, the equivalence principle is sacrificed in such a way that we can not trust the usual effective field theory in the zone. Adopting entanglement with the early radiation, we can no longer afford entanglement between the two halves of the near-horizon region which translates into presence of energetic quanta around the horizon, or a firewall. 

A natural attempt to fully identify the interior with the distant radiation is not only extremely non-local but also leads to the frozen vacuum \cite{frozen} or inability of the infalling observer to excite the near-horizon state which violates the equivalence principle no less than a firewall.

A possible alternative would be to resort to strong complementarity by arguing that it is not a problem when two observers see absolutely different physical content of the zone if they cannot communicate their findings to each other. It does not seem to work out in an ideal way because an infalling observer might perform a precise measurement of the early radiation before entering the zone, or because there is some time for a freely falling observer to change his mind and turn around inside the zone before crossing the horizon \cite{Bousso}. However, the contradictory measurements and inferences might turn to be computationally unfeasible \cite{HH}, extremely fine-grained \cite{Susskind2}, overwhelmingly affecting the Black Hole state, or even be akin to observing quantum superpositions of macroscopic worlds \cite{NV}. 

The problem remains controversial, and it is only clear that new insights are needed and, hopefully, expected. They would certainly deepen our understanding of quantum mechanics and gravity; and the relevant issues are really exciting.  For example, one line of reasoning \cite{SusskindRev} asserts that the very measurement made by the distant observer creates high energy quanta, or a firewall, which would kill the infalling colleague. If this process is to be causally conceivable, and if we do insist on causality of so violent behaviours, then Einstein-Rosen bridges between the Black Hole and its early radiation must be invoked \cite{MaldacenaSusskind}.

\vspace{0.5cm}

Instead, we would like to offer a different approach to the AMPS problem which might actually point at better integrity of physical description. We argue that the paradoxes might be resolved by taking quantum gravity effects into account in the form of unavoidable entanglement with microscopic geometrical configurations of spacetime. 

Our main idea is that locally an effective field theory description can be valid and very precise everywhere in the low curvature regions, although taken all the way over huge spacetime distances, the tiny errors might accumulate considerably enough to entail the loss of purity of the early Hawking radiation. Note, for an illustration, that mean calendar year length difference of Gregorian and Julian calendars is less than eleven minutes and at first glance seems impractical, but in four hundred years it sums up to three full days. Random errors do not grow as fast as a gradual change but still can eventually matter. 

Below we give some simple estimates of quantum gravity effects on propagating radiation, and show how the Page and scrambling times can naturally appear from such considerations.

\vspace{0.5cm}

Let us first address the clear-cut problem at the Page time $t_{\mathrm P}\propto M^3$. If we want to take the quantum gravity effects into account, then probably it would be safest and fairly model-independent to assume that the wavelength of a typical photon of Hawking radiation $$\lambda\sim \frac{M}{m_{Pl}}\cdot l_{Pl}$$ cannot be determined with precision better than the Planck length $l_{Pl}$. We treat it as an intrinsic fluctuation of the wavelength $$\delta\lambda\sim l_{Pl}.$$ If the photon has propagated over a huge number $N$ of wavelengths, then the statistical uncertainty of the path length $L=N\lambda$ amounts to $$\delta L\sim\sqrt{N}\cdot l_{Pl}$$ from which we see that $\delta L$ reaches $\lambda$ when $$L\sim \frac{\lambda^3}{l^2_{Pl}}\sim \left(\frac{M}{m_{Pl}}\right)^3\cdot l_{Pl}\sim ct_{\mathrm P}.$$ 

Therefore, to the Page time, the information about the relative phases of different photons is definitely lost. Of course, the reason is that we treat the geometry as a mute background arena for electrodynamics. But the actual quantum state of photons gets dynamically entangled with quantum fluctuations of geometry and, after a long enough time, tracing over the states of geometry produces a very blurred image of the emitted light. Since it is no longer pure, there are no obstructions for the late radiation to be entangled with the Black Hole interiors. Quantum gravity has gone into full play despite the incredible smallness of all its local effects.

\vspace{0.5cm}

The scrambling time $t_{scr}\propto M\log M$ is much trickier to discuss in this context. But we argue that a more careful treatment of the decoherence features allows to naturally come at this time scale, too. Let us assume that the dynamics of photons in presence of small quantum gravitational corrections can be described as an open quantum system with the Lindblad equation \cite{Lindblad,Kossakowski} $$\dot\rho= {\hat{\mathcal L}}\rho$$ for the density matrix $\rho$. It provides a natural framework for discussing phenomenology of quantum gravity, see for example the paper \cite{Andrianov} and references therein for possible effects in the oscillations of neutral kaons. The dynamical semigroup generator ${\hat{\mathcal L}}$ consists of two terms: the commutator with the Hamiltonian which reproduces the standard Schr{\" o}dinger equation and the additional Lindblad operator, the required properties of which we do not need to discuss now. 

The order of magnitude of the coefficients in the matrix of the Lindblad operator depends on the adopted level of coarse graining. We want to address situations in which an effective field theory description is just marginally enough to come to a contradiction. It is natural then to adopt a resolution at the level of ultraviolet cutoff scale which is presumably the Planck scale. According to our considerations above, we can expect to start confusing the neighbouring states after the time period $\tau\sim\frac{\lambda}{c}$, and therefore the matrix entries of the Lindblad operator are expected of order $\frac{c}{\lambda}\sim\frac{m_{Pl}}{M t_{Pl}}$. 

Note then that we are not interested in simply converting one state into another with off-diagonal elements of the Lindblad operator because it does not automatically entail decoherence, even though it may produce very interesting effects such as CPT-violation \cite{Andrianov}. We would rather like to find an independent growth of probabilities for other states bringing the system to a statistical mixture. For the initial Cauchy data, we can assume that a given photon has been in a given pure state with the probability $p_1$ practically equal to $1$. However, due to quantum gravity effects, the other states could not have been totally absent. Their initial probabilities $p_j (0)$ can be estimated as $\sim\frac{m_{Pl}}{M}$, or some mild power of it. We see that typically $\log p_j$ starts growing with time as $\frac{m_{Pl}}{M t_{Pl}}\cdot t$. And extrapolating this trend far beyond any reasonable limit, an undoubtedly mixed state is reached when $$\log \frac{M}{m_{Pl}}\sim \frac{m_{Pl}}{M t_{Pl}}\cdot t,$$ or at the fast scrambling time. Admittedly, it sounds rather speculative, but so is the issue of scrambling time itself in the context of firewalls.

\vspace{0.5cm}

Of course, it was always clear that somehow relaxing the assumption of entanglement between the early and the late radiation would give a way to resolving the paradox \cite{Sabine}. Moreover, a concrete realisation was proposed in the context of Many Worlds interpretation of quantum mechanics \cite{Hsu}. A Black Hole randomly emits really huge amounts of low-energy quanta, and therefore, for an old specie, its position must be very indefinite due to recoil effects \cite{Page}. Accordingly, we have to face macroscopic superpositions in the system. (We note in passing, it might be interesting to compare these superpositions with those which appear in the Ref. \cite{NV}.) It is certainly a logical possibility that after specifying a certain branch for the macroscopic world the unitarity is lost despite being safe in the full picture \cite{Hsu2}. However, it seems rather radical an idea which probably could make macroscopic quantum superpositions (too easily) observable. 

Our proposal is different. We talk about small fluctuations of geometry irrespective to the foundational issues of quantum mechanics. Note also that what we mean is not just gravitons radiated from the Black Hole which are of little or no interest for resolving the paradox, but it is really an effect of a slightly random medium with the spacetime foam on the way of the photons. Of course, it would be completely legitimate to wonder how the distant fluctuations and the nearby emission effects contrive to avoid the potential tensions and save the day. 

Unfortunately, we are not so much aware of the details of quantum gravity and even its real degree of non-locality. However, our point is that these non-local quantum gravity effects need not be locally observable with any deviations from effective field theory or with other conceivable types of anomalies such as the frozen states. What matters is only the Planck scale physics.

Although we are very far from giving the final and definitive solutions, the estimates look very interesting because they produce the relevant time scales from a completely different side. Actually, it is not the first time when Black Holes teach us non-trivial lessons about the Nature. Very remarkably, Black Holes obey the usual  laws of thermodynamics \cite{Bekenstein} which should not be expected of a simple and fairly isolated system. Somehow, general relativity has given the hints to a deeper parent theory which has to describe Black Holes as statistical systems with many degrees of freedom. And now we might learn some new lessons about quantum gravity regimes from an unexpected direction. We ought to be ready and open-minded for new insights. Black Holes have a good credit history, and it would be a nice idea to take seriously what they say.

\vspace{0.5cm}

{\bf Acknowledgements.} The author is grateful to A. Andrianov for useful discussions. The author was partially supported by Saint Petersburg State University research grant No. 11.38.660.2013 and by Russian Foundation for Basic Research grant No. 12-02-31214.

\end{document}